\documentclass[useAMS,usenatbib,usegraphicx]{mn2e}
\usepackage{psfig}   
\usepackage{graphicx}
\usepackage{subfigure}

\title[Dynamics of stars and brown dwarfs]{Spatial differences between stars and brown dwarfs: a dynamical origin?}

\author[R.~J.~Parker \& M.~Andersen]{
  Richard J.~Parker$^{1}$\thanks{E-mail: rparker@phys.ethz.ch} and Morten Andersen$^2$
  \vspace*{0.1cm}\\
   $^1$ Institute for Astronomy, ETH Z{\"u}rich, Wolfgang-Pauli-Strasse 27, 8093, Z{\"u}rich, Switzerland \\
   $^2$ Institut de Plan{\'e}tologie et d'Astrophysique de Grenoble, BP 53, F-38041 Grenoble C{\'e}dex 9, France}

\begin{document}

\date{Accepted for publication in MNRAS}
                             
\pagerange{\pageref{firstpage}--\pageref{lastpage}} \pubyear{2014}

\maketitle

\label{firstpage}

\begin{abstract}
We use $N$-body simulations to compare the evolution of spatial distributions of stars and brown dwarfs in young star-forming regions. We use three different diagnostics; the ratio of stars to 
brown dwarfs as a function of distance from the region's centre, $\mathcal{R}_{\rm SSR}$, the local surface density of stars compared to brown dwarfs, $\Sigma_{\rm LDR}$, and we compare the global spatial 
distributions using the $\Lambda_{\rm MSR}$ method. From a suite of twenty initially statistically identical simulations, 6/20 attain $\mathcal{R}_{\rm SSR} << 1$ \emph{and}  $\Sigma_{\rm LDR} << 1$  \emph{and} $\Lambda_{\rm MSR} << 1$, 
indicating that dynamical interactions could be responsible for observed differences in the spatial distributions of stars and brown dwarfs in star-forming regions. However, many simulations also display apparently contradictory 
results -- for example, in some cases the brown dwarfs have much lower local densities than stars ($\Sigma_{\rm LDR} << 1$), but their global spatial distributions are indistinguishable ($\Lambda_{\rm MSR} = 1$) and the relative proportion 
of stars and brown dwarfs remains constant across the region ($\mathcal{R}_{\rm SSR} = 1$).  Our results suggest that extreme caution should be exercised when interpreting any observed difference in the spatial distribution 
of stars and brown dwarfs, and that a much larger observational sample of regions/clusters (with complete mass functions) is necessary to investigate whether or not brown dwarfs form through similar mechanisms to stars.  
\end{abstract}


\section{Introduction}

One of the outstanding questions in star formation is whether the mechanism through which brown dwarfs (objects not massive enough to burn hydrogen in their cores) form is more like that of higher (e.g. Solar) mass stars, or 
more like that of giant planets. This can be addressed by comparing the various properties of brown dwarfs (BDs) with stars, such as multiplicity \citep{Duchene13b}, kinematics \citep{Luhman07} and spatial distribution \citep{Kumar07}. 

Several studies \citep[e.g.][]{Luhman06,Bayo11,Parker11b,Parker12c} have shown that BDs have a similar spatial distribution to stars in some star-forming regions; but there are other regions where the BDs appear to be more spread out \citep{Kumar07,Caballero08,Kirk12}. 
Furthermore, several studies \citep{Andersen11,Suenaga13} have determined the ratio of stars to BDs (the `substellar ratio' $\mathcal{R}_{\rm ss}$) as a function of distance from the centre of the Orion Nebular Cluster (ONC) and there is tentative evidence for 
a decrease in $\mathcal{R}_{\rm ss}$ as a function of distance from the cluster centre, though measuring the substellar mass function in this region (and others) remains challenging \citep[e.g.][]{Alves12,DaRio12,Lodieu12}.

Taken at face value, these results suggest that brown dwarfs have different spatial distributions to stars in some (but not all) star forming regions and clusters. This could imply that brown dwarfs form through a different mechanism to stars in those regions, 
\citep[e.g.][]{Thies08}, or perhaps that dynamical interactions alter their spatial distribution in some regions \citep[e.g.][]{Adams02,Reipurth01,Goodwin05c}, but not others. In order to test this, $N$-body simulations (which can be repeated many times with different random number seeds 
to guage the level of stochasticity in the initial conditions) 
of the evolution of young star forming regions should be analysed with the same method(s)/techniques(s) used to analyse observational data.

In this paper, we use three different diagnostics to compare the spatial distributions of stars and BDs in numerical simulations of the evolution of star-forming regions. We measure the ratio of stars to BDs ($\mathcal{R}_{\rm ss}$) as a function of distance from the cluster centre; 
we compare the `local density ratio' of stars and BDs using the $\Sigma_{\rm LDR}$ method \citep{Maschberger11,Parker14b}, and we compare the global spatial distributions using the `mass segregation ratio' $\Lambda_{\rm MSR}$ \citep{Allison09a}.
We then re-examine the ONC data from \citet{Andersen11} to look for differences in the local density of BDs compared to stars using $\Sigma_{\rm LDR}$, and the relative spatial distribution using $\Lambda_{\rm MSR}$.

\section{Quantifying differences between stars and brown dwarfs}

The ratio of stars to brown dwarfs, the `substellar ratio' $\mathcal{R}_{ss}$ has been measured in several star-forming regions and the field \citep[e.g.][]{Briceno02,Luhman04,Guieu06,Andersen08,Andersen11,Scholz12,Suenaga13}. Often, the global $\mathcal{R}_{ss}$ is compared between different regions 
to search for environmental dependencies \citep[e.g.][]{Scholz12} but \citet{Andersen11} also measure  $\mathcal{R}_{ss}$ 
as a function of distance from the centre of the Orion Nebula Cluster (ONC), and find that it decreases so that the ratio of the outer bin $\mathcal{R}_{ss}$ to inner bin $\mathcal{R}_{ss}$:
\begin{equation}
\mathcal{R}_{\rm SSR} = \mathcal{R}_{ss, out}/\mathcal{R}_{ss, in},
\label{rss}
\end{equation}
is significantly less than unity (in that there is a $\sim$1.5-$\sigma$ difference between the observed inner and outer values). 

The `mass segregation ratio', $\Lambda_{\rm MSR}$ \citep{Allison09a} determines the level of mass segregation based on the length of the minimum spanning tree (MST) of a chosen subset of $N_{\rm MST}$ objects in the region $l_{\rm subset}$, compared to the average length of the minimum spanning tree of many randomly drawn $N_{\rm MST}$ objects, 
$\langle l_{\rm average} \rangle$, with the lower (upper) uncertainty taken to be the MST length which lies 1/6 (5/6) of the way through an ordered list of all the random lengths  (${\sigma_{\rm 1/6}}/{l_{\rm BDs}}$ or ${\sigma_{\rm 5/6}}/{l_{\rm BDs}}$). 
In this paper, we will compare the MSTs of brown dwarfs to the cluster average:
\begin{equation}
\Lambda_{\rm MSR} = {\frac{\langle l_{\rm average} \rangle}{l_{\rm BDs}}} ^{+ {\sigma_{\rm 5/6}}/{l_{\rm BDs}}}_{- {\sigma_{\rm 1/6}}/{l_{\rm BDs}}}.
\label{lambda}
\end{equation}
Thus far, the $\Lambda_{\rm MSR}$ method has only been applied to two observed star-forming regions to look for differences in the spatial distribution of BDs compared to stars; in both Taurus \citep{Parker11b} and $\rho$~Oph \citep{Parker12c} the BDs have the same spatial 
distribution as the stars.

The `local surface density ratio', $\Sigma_{\rm LDR}$ compares the median local surface density of a chosen subset of stars to the median value of either the entire region, or another chosen subset \citep{Maschberger11,Kupper11,Parker14b}. The surface density, $\Sigma$, is determined 
as in \citet{Casertano85}:
\begin{equation}
\Sigma = \frac{N - 1}{\pi r_N^2},
\end{equation}
where $r_N$ is the distance to the $N^{\rm th}$ nearest star and we adopt $N = 10$ throughout this work.

In this paper, we compare the brown dwarfs to all stars with mass $m < 1$M$_\odot$:
\begin{equation}
\Sigma_{\rm LDR} = \frac{\tilde{\Sigma}_{\rm BDs}}{\tilde{\Sigma}_{0.08 \leq m/{\rm M_\odot} < 1.0}}, 
\label{sigma}
\end{equation}
and use the two-dimensional Kolmogorov-Smirnoff (KS) test from \citet{Press92} to determine whether or not two subsets can share the same parent distribution. If $\Sigma_{\rm LDR} < 1$ and the calculated KS p-value is lower than 0.1, then we consider the 
local density of brown dwarfs to be significantly lower compared to stars. 
Using $\Sigma_{\rm LDR}$, \citet{Parker12c} found no evidence for systematically different local densities of BDs compared to stars in $\rho$~Oph. \citet{Kirk12} used a variation of $\Sigma_{\rm LDR}$ and found that low-mass stars and BDs typically have lower surface densities than higher mass stars in the 
Gomez groups in Taurus, IC~348 and the ONC, but not in Chamaeleon~I or Lupus.

\section{$N$-body simulations}
\label{results}

\subsection{Initial Conditions}

In the following analysis, we use only one set of initial conditions for star forming regions, which we deem to be the most dynamically extreme in terms of the number of ejections of, and the maximum density experienced  by, the stars and brown dwarfs \citep{Allison12}. 

The star-forming 
regions consist of 1500 objects, distributed randomly in a fractal with dimension $D = 1.6$ and radius $r_F = 1$\,pc. This fractal dimension results in a very clumpy distribution, which can lead to the ejection of low-mass objects from the clumps. However, the initial 
spatial distributions of stars and BDs are indistinguishable. The global virial ratio (defined as $\alpha_{\rm vir} = T/|\Omega|$, where $T$ and $|\Omega|$ are the total kinetic energy and
total potential energy of the stars, respectively) is $\alpha_{\rm vir} = 0.3$, i.e.\,\,subvirial. For the exact details of the spatial set-up, and the velocity distribution of stars and brown dwarfs, we refer the interested reader to \citet{Goodwin04a} and \citet{Parker14b}. 

We draw primary masses from the \citet{Maschberger13} formulation of the IMF. We then assign binary separations based on the primary mass \citep[the mean separation decreases with decreasing primary mass,][]{Burgasser07,Raghavan10,Bergfors10,Janson12,Sana13,deRosa14} and 
mass ratios drawn from a flat distribution \citep{Metchev09,Reggiani11a,Reggiani13,Duchene13a}. 
Finally, eccentricities are drawn from  a flat distribution \citep{Abt06,Raghavan10}. This set-up results in a global \emph{system} star-to-brown-dwarf ratio of 4:1, consistent with both the Galactic field and star-forming regions \citep{Chabrier05,Andersen08,Bochanski10}.

We evolve the star forming regions for 10\,Myr using the \texttt{KIRA} integrator in the \texttt{STARLAB} package \citep{Zwart99,Zwart01}. We do not include stellar evolution in the simulations.

\subsection{Dynamical evolution over 10\,Myr}


\begin{figure*}
  \begin{center}
\setlength{\subfigcapskip}{10pt}
\vspace*{-0.3cm}
\hspace*{-1.cm}\subfigure[$\mathcal{R}_{ss, out}/\mathcal{R}_{ss, in}$]{\label{lines-a}\rotatebox{270}{\includegraphics[scale=0.27]{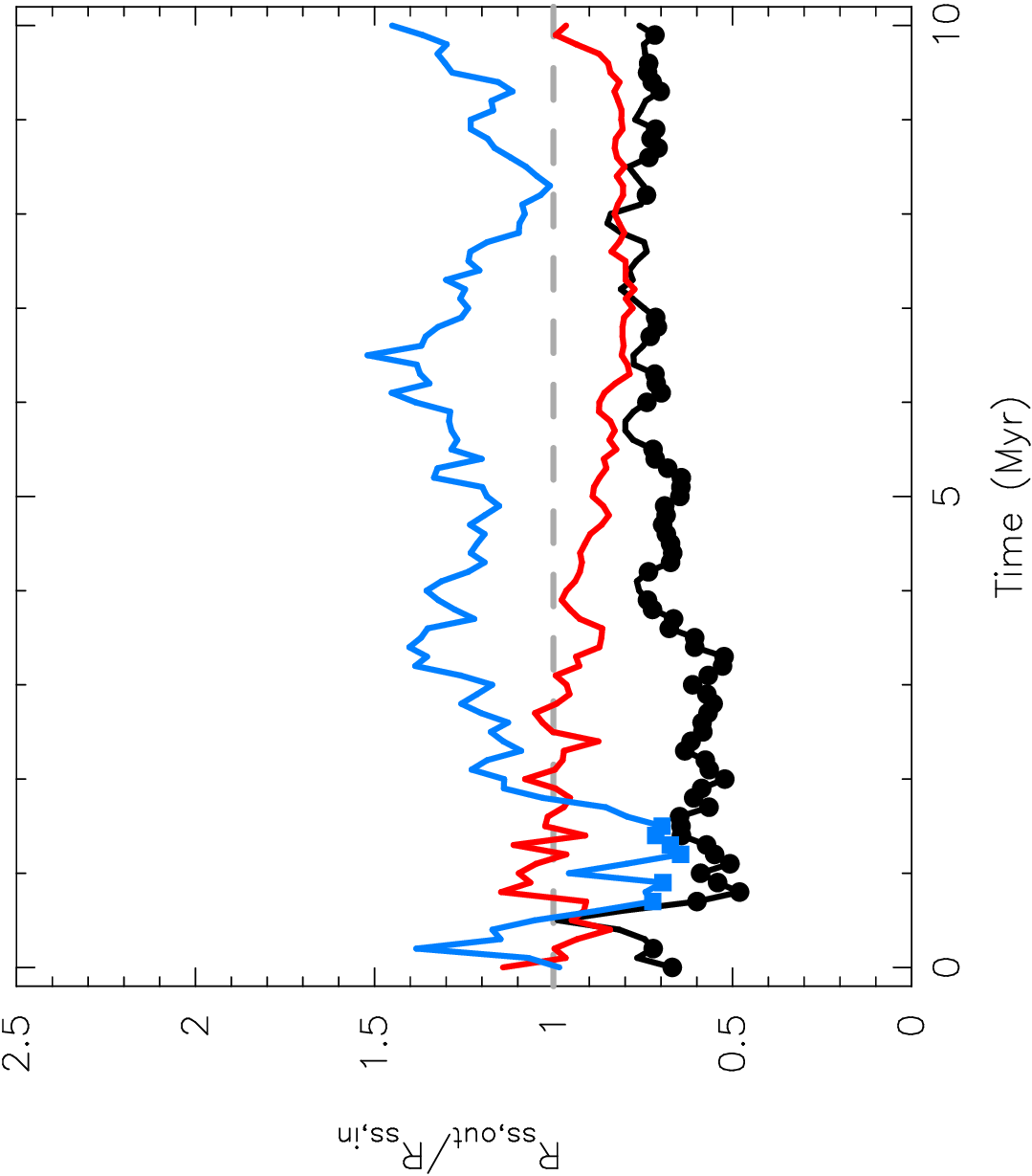}}}
\hspace*{0.3cm} 
\subfigure[$\Lambda_{\rm MSR}$ (BDs)]{\label{lines-b}\rotatebox{270}{\includegraphics[scale=0.27]{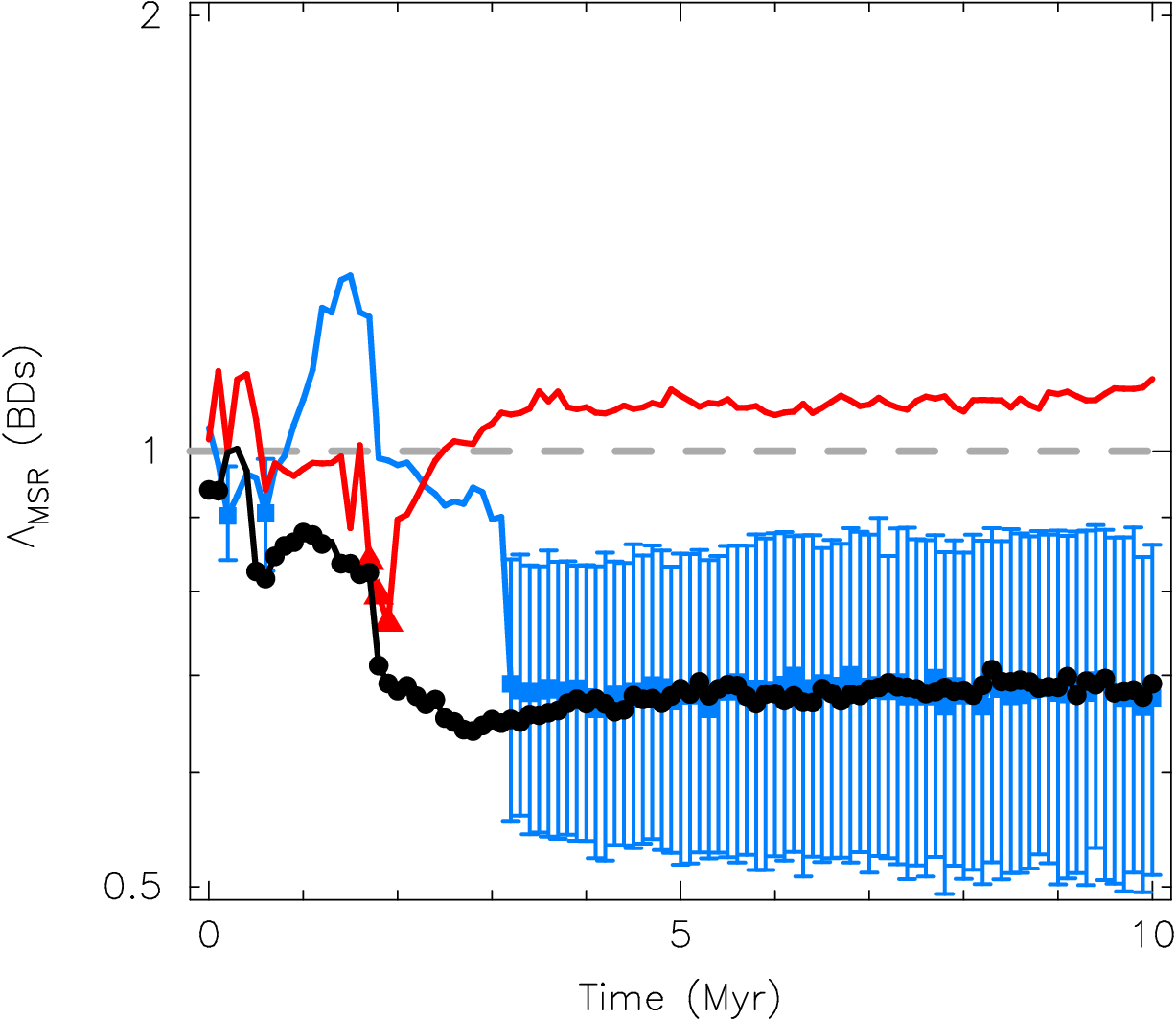}}}
\hspace*{0.3cm} 
\subfigure[$\Sigma_{\rm LDR}$ (BDs/stars)]{\label{lines-c}\rotatebox{270}{\includegraphics[scale=0.27]{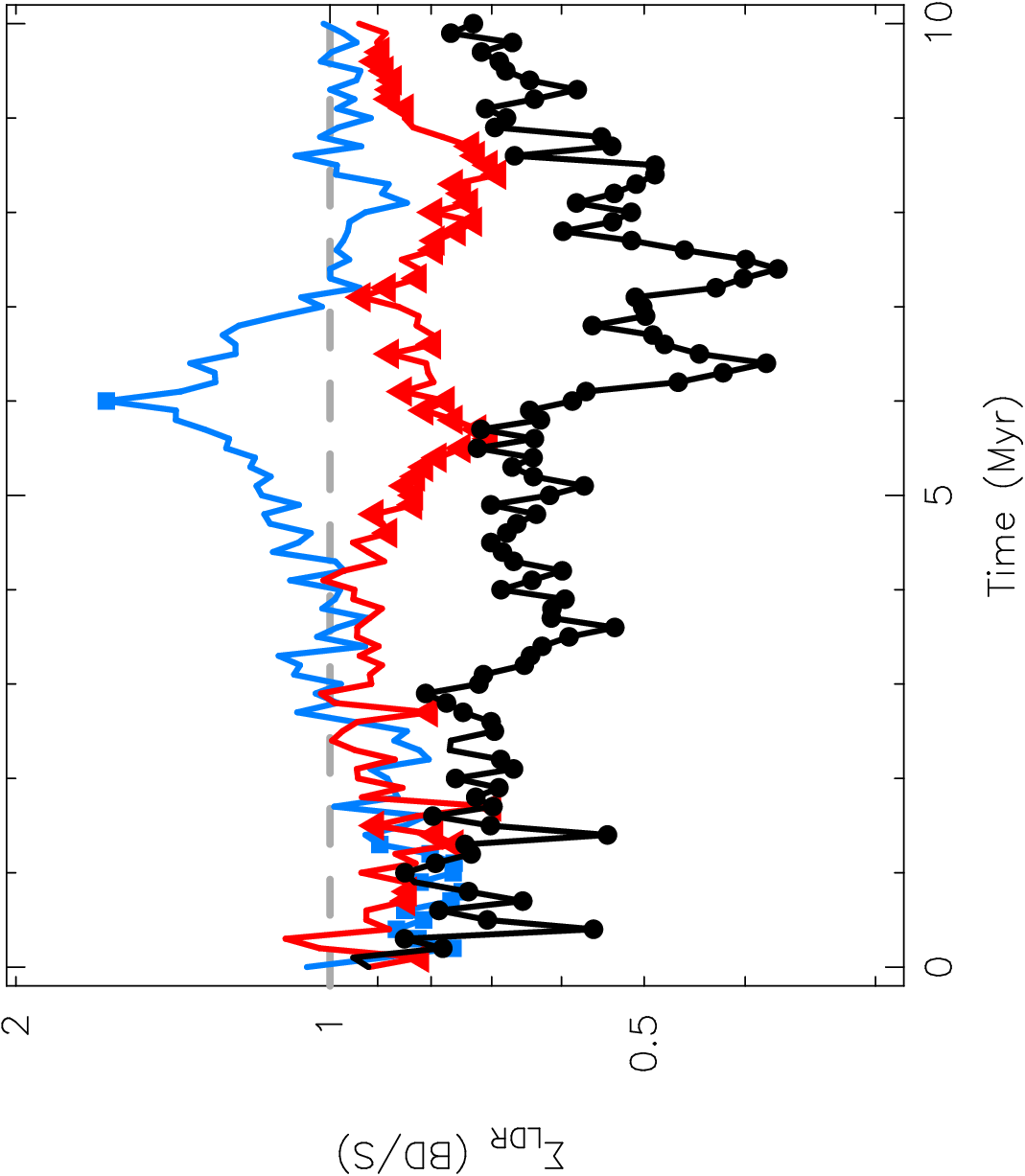}}}
\caption[bf]{Evolution of $\mathcal{R}_{ss, out}/\mathcal{R}_{ss, in}$, $\Lambda_{\rm MSR}$ and $\Sigma_{\rm LDR}$ with time for three individual simulations. For each measurement, we plot a filled symbol when the value signficantly deviates from unity. In panel (b) we show the uncertainties on the $\Lambda_{\rm MSR}$ 
measurements according to Eq.~\ref{lambda} for one simulation. $\mathcal{R}_{ss, out}/\mathcal{R}_{ss, in} < 1$ indicates that more brown dwarfs are located 
on the outskirts of the cluster than in the centre, $\Lambda_{\rm MSR} < 1$ indicates that the brown dwarfs collectively have a more sparse spatial distribution, and $\Sigma_{\rm LDR} < 1$ indicates that the brown dwarfs have lower local density than more massive objects.}
\label{lines}
  \end{center}
\end{figure*}

The evolution of the star forming regions follow the same qualitative
pattern; substructure is erased within the first $\sim$1\,Myr \citep{Goodwin04a,Allison10,Parker12d} and the subvirial velocities lead to violent relaxation and collapse to a centrally 
concentrated, bound cluster \citep{Parker12d,Parker14b}. The adopted initial conditions lead to an ejected halo of objects on the outskirts of the cluster \citep{Allison12}. However, the 
evolution of other parameters is highly stochastic; some clusters exhibit mass segregation whereas others do not \citep{Allison10,Parker14b}, and the binary population (both stars and 
brown dwarfs) can be altered to varying degrees \citep{Parker12b}.

Because the cluster expands due to two-body interactions \citep{Moeckel12,Gieles12,Parker12d} it is difficult to define a radially varying $\mathcal{R}_{\rm ss}$ ratio for annuli of fixed 
physical width. For this reason we adopt four annuli from the cluster centre-of-mass; $0 - 0.25$\,$r_c$; $0.25 - 0.50$\,$r_c$; $0.50 - 0.75$\,$r_c$; and $0.75 - 0.95$\,$r_c$, where $r_c$ is the total extent of the cluster in the $N$-body simulation. 
We exclude the very outskirts ($>$95\,per cent) of the cluster -- i.e.\,\,ejected stars, though we note that in future the \emph{Gaia} satellite may be able to trace the birth-sites of ejected BDs from clusters. 
We then compute the $\mathcal{R}_{\rm SSR} = \mathcal{R}_{ss, out}/\mathcal{R}_{ss, in}$ ratio as the ratio of the outer annulus to the inner.

We determine $\Lambda_{\rm MSR}$ for the 2D distribution within 95\,per cent of the cluster centre at each simulation snapshot and compare the MST of the 50 lowest mass ($<$0.02\,M$_\odot$)
objects to randomly chosen MST lengths. We choose 50 objects to strike a balance between having too few links in the MST (which would produce a very noisy signal), and too many (which would be washed out against the mean MST). 
We also determine the local density ratio $\Sigma_{\rm LDR}$ for all brown dwarfs, compared to stars with masses less than 1\,M$_\odot$, 
again in two dimensions within 95\,per cent of the cluster members.

\begin{figure*}
  \begin{center}
\setlength{\subfigcapskip}{10pt}
\vspace*{-0.3cm}
\hspace*{-1.cm}\subfigure[$\mathcal{R}_{ss, out}/\mathcal{R}_{ss, in}$]{\label{med-a}\rotatebox{270}{\includegraphics[scale=0.27]{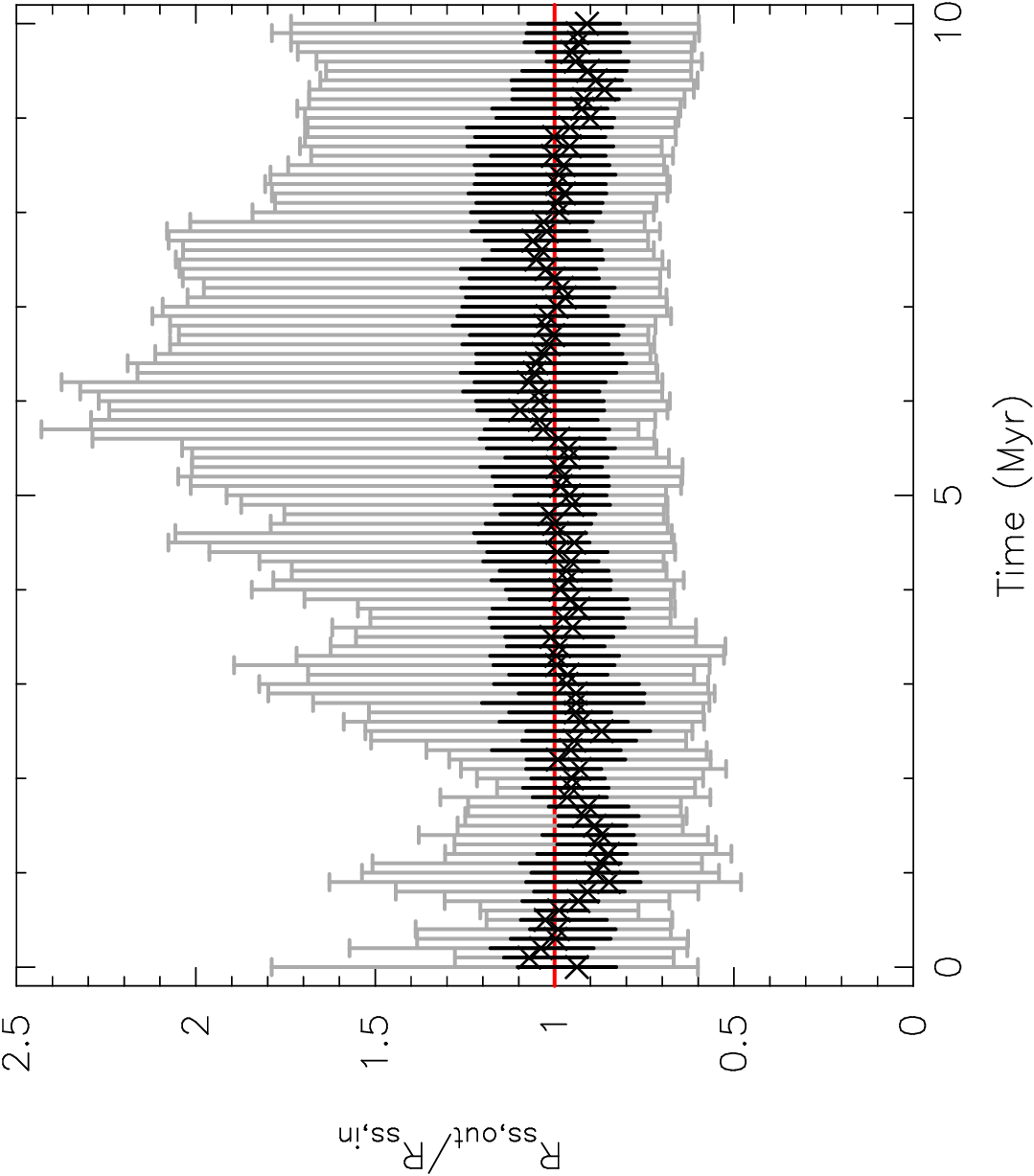}}}
\hspace*{0.3cm} 
\subfigure[$\Lambda_{\rm MSR}$ (BDs)]{\label{med-b}\rotatebox{270}{\includegraphics[scale=0.27]{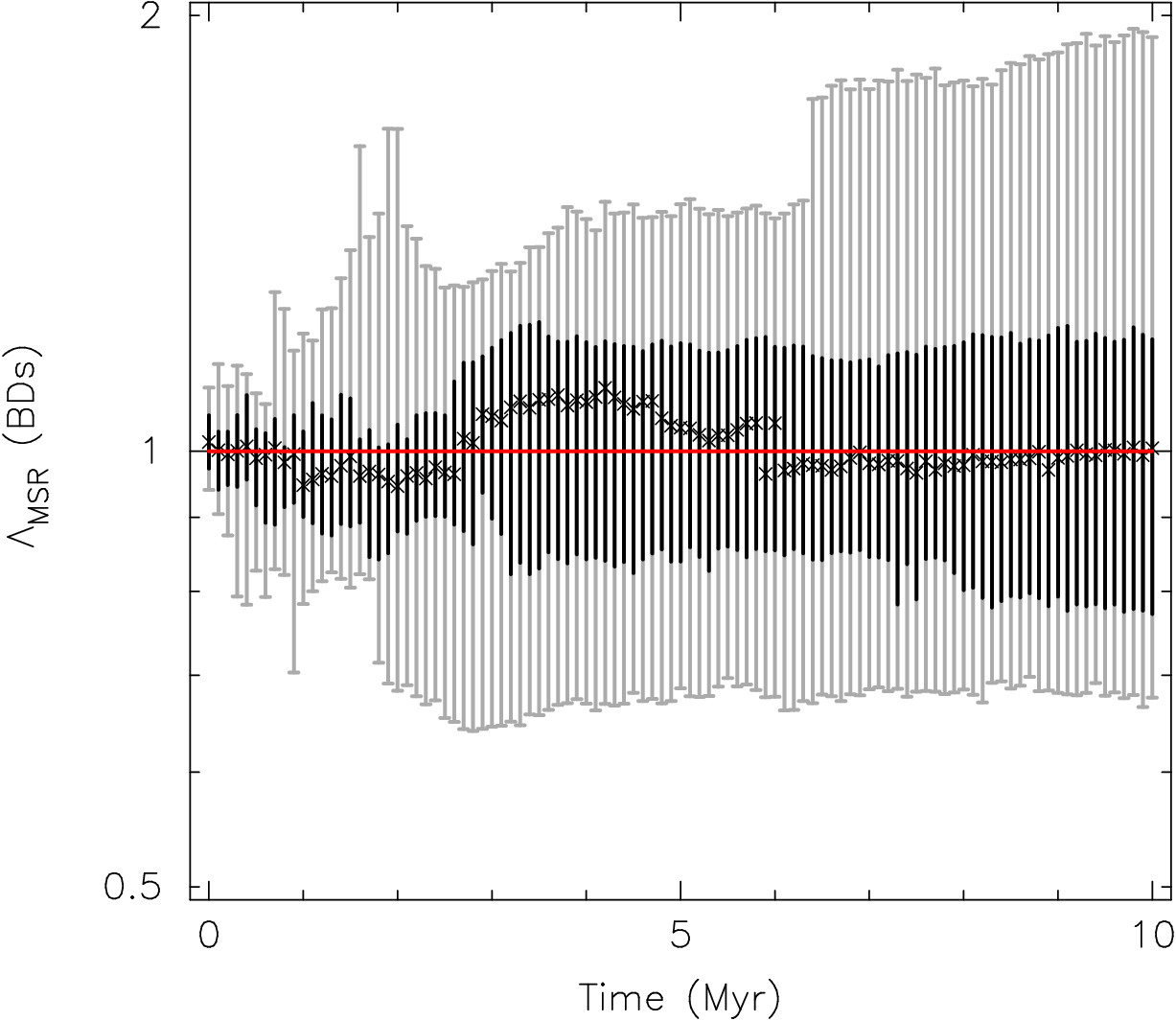}}}
\hspace*{0.3cm} 
\subfigure[$\Sigma_{\rm LDR}$ (BDs/stars)]{\label{med-c}\rotatebox{270}{\includegraphics[scale=0.27]{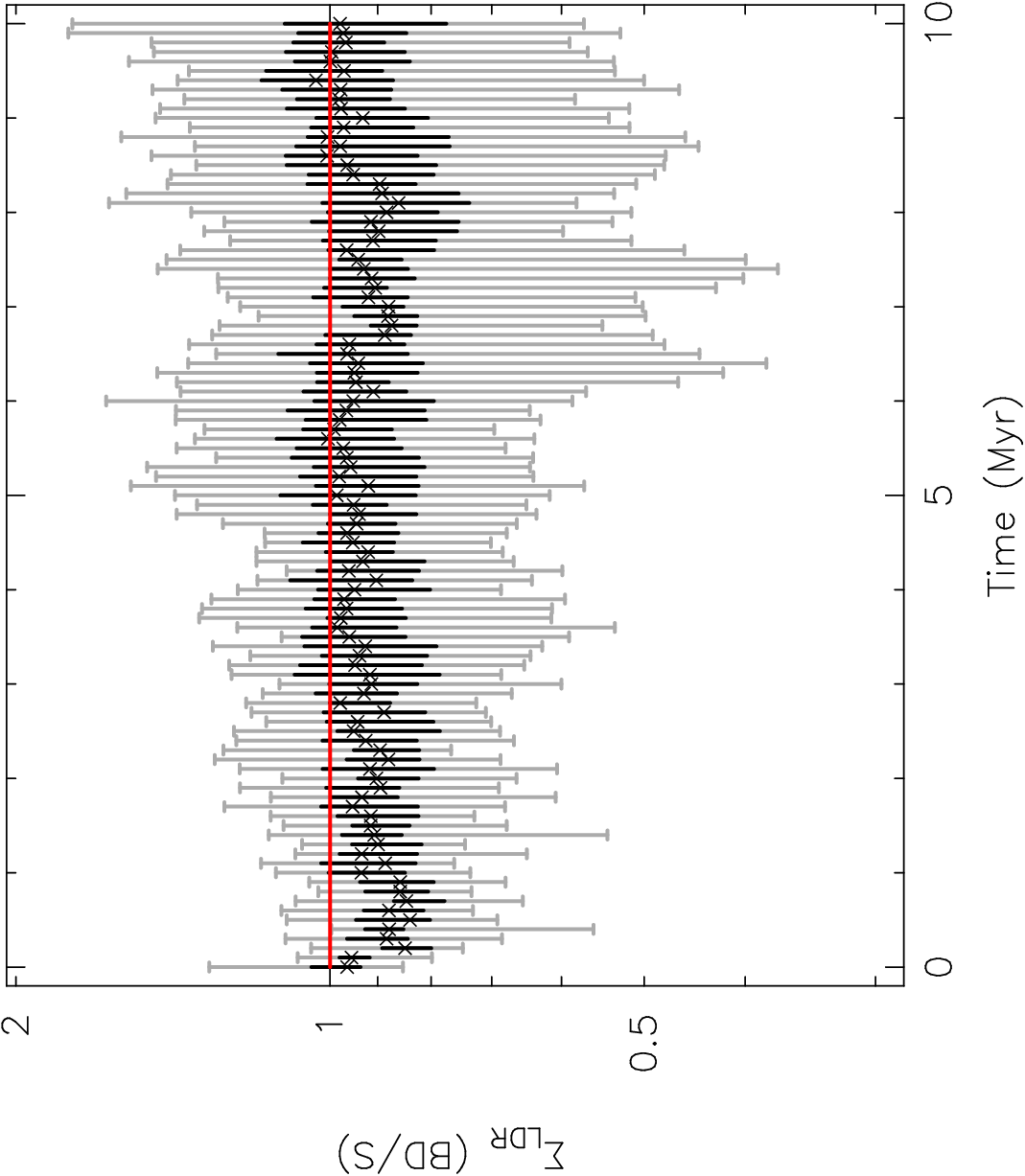}}}
\caption[bf]{Evolution of $\mathcal{R}_{ss, out}/\mathcal{R}_{ss, in}$, $\Lambda_{\rm MSR}$  and $\Sigma_{\rm LDR}$ (BDs/stars) for all 20 simulations. 
Each panel shows the median value of 20 simulations with identical initial conditions (the crosses) and the darker `error bars' indicate 
25 and 75 percentile values. The entire range of possible values from the 20 sets of initial conditions is shown by the lighter `error bars'. }
\label{med}
  \end{center}
\end{figure*}

 We use the 95\,per cent extent and perform our calculations in 2D to attempt to mimic the information available to observers. However, we also repeated the analysis in 3D for stars which are energetically bound to the cluster using the method outlined in \citet{Baumgardt02}, 
and in a very conservative calculation we repeated the original 2D determination but limited the extent to 85\,per cent of the cluster. Both of these alternative determinations give very similar results to our default calculation.

In Fig.~\ref{lines} we show the evolution of $\mathcal{R}_{\rm SSR}$, $\Lambda_{\rm MSR}$ and $\Sigma_{\rm LDR}$ for three out of our suite of 20 simulations. In each panel, we plot a filled 
symbol when the deviation from unity is significant (more than 2-$\sigma$) for each measure. In panel (b) we show the uncertainty associated with $\Lambda_{\rm MSR}$ as defined 
by Eq.~\ref{lambda} for one simulation -- the uncertainties on the remaining simulations are not shown because the plot would become unreadable, but are similar in size. The magnitude of the uncertainties associated with 
$\mathcal{R}_{\rm SSR}$ and $\Sigma_{\rm LDR}$ are also comparable.

In the first simulation (the black lines/circles), the $\mathcal{R}_{\rm SSR}$ ratio is actually signficantly less than 
unity before dynamical evolution occurs (despite their spatial distributions being the same). This ratio rises to unity during the cool collapse, but then is significantly less than unity for the remainder of the simulation. This could 
be interpreted as the brown dwarfs being ejected into the outskirts of the cluster, and if this is the case we might expect them to have a more sparse spatial distribution than the 
stars. This is confirmed by the $\Lambda_{\rm MSR}$ ratio, which shows the brown dwarfs to be more spatially spread out with respect to the average cluster members. Furthermore, 
the $\Sigma_{\rm LDR}$ ratio shows that on average, the local surface density around brown dwarfs to be lower than for stars. Taken together, the natural interpretation is that dynamical
 interactions have ejected the brown dwarfs to the cluster periphery.

If we examine each simulation individually, we find that at various points in the whole 10\,Myr of evolution, 6/20 simulations have $\mathcal{R}_{\rm SSR} << 1$ \emph{and} $\Lambda_{\rm MSR} << 1$ 
\emph{and} $\Sigma_{\rm LDR} <<1$. The simulation shown by the black points/lines in Fig.~\ref{lines} displays significant differences between the spatial distributions of stars and brown dwarfs in all three diagnostics for a total of 2.8\,Myr, and significant 
differences in two of the three diagnostics for another 7.0\,Myr in total. There are another five simulations which show differences in all three diagnostics, but for a much shorter total time: 0.4, 0.3, 0.1, 0.1 and 0.1\,Myr. 14/20 simulations 
show significant differences in two of three diagnostics for some of their evolution (the median length is 0.5\,Myr), and \emph{all} simulations show a difference between the spatial distributions of stars and brown dwarfs in at least one diagnostic for some of their evolution (the median length is 2.7\,Myr).  

However, if we examine another simulation (the blue lines/squares) we see that the $\mathcal{R}_{\rm SSR}$ ratio is significantly lower than unity in the first 2\,Myr, before becoming 
more than unity (i.e.\,\,there are relatively more brown dwarfs than stars in the central region, compared to the outskirts). At the same time, $\Lambda_{\rm MSR}$ suggests that the brown dwarfs are more spread 
out from 3\,Myr onwards, whereas  $\Sigma_{\rm LDR}$ indicates that the BDs are not in regions of lower local density than the stars. In a third simulation (the red lines/triangles) neither 
$\mathcal{R}_{\rm SSR}$ nor $\Lambda_{\rm MSR}$  are significant, yet the $\Sigma_{\rm LDR}$ ratio taken in isolation would suggest that the BDs are in locations of lower surface density 
than the stars. 

In order to guage the significance of these particular simulations, we plot the evolution of each of our chosen metrics for all 20 simulations in Fig.~\ref{med}. The crosses indicate 
the median value from 20 simulations at each snapshot, whereas the black `error bars' indicate the 25 and 75 percentiles, and the full range in the simulations is shown by the 
grey `error bars'. (Note that these are not error bars in the conventional sense -- we are only showing the range of values from 20 simulations at a given time, and not the uncertainty on the measurement.) 
On average, each measurement does not significantly deviate from unity, suggesting that dynamical processing cannot be the mechanism which results in 
different spatial distributions of brown dwarfs compared to stars. However, as we have seen, using only one metric can lead to erroneous (or at the very least na{\"i}ve) conclusions.

\section{Data for the ONC}
\label{method}

\begin{figure*}
  \begin{center}
\setlength{\subfigcapskip}{10pt}
\vspace*{-0.3cm}
\hspace*{-1.cm}\subfigure[]{\label{test-a}\rotatebox{270}{\includegraphics[scale=0.35]{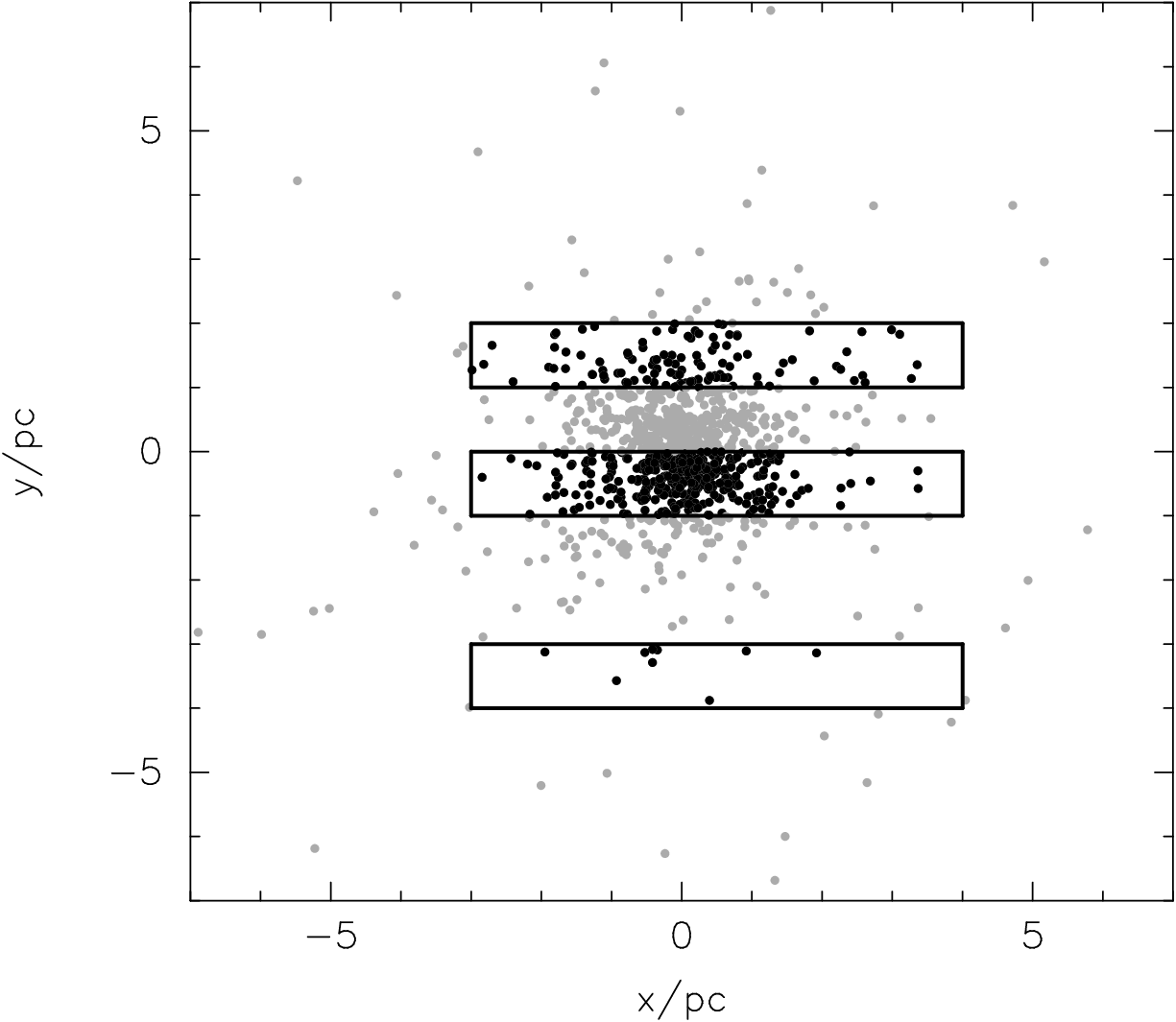}}}
\hspace*{0.3cm} 
\subfigure[]{\label{test-b}\rotatebox{270}{\includegraphics[scale=0.37]{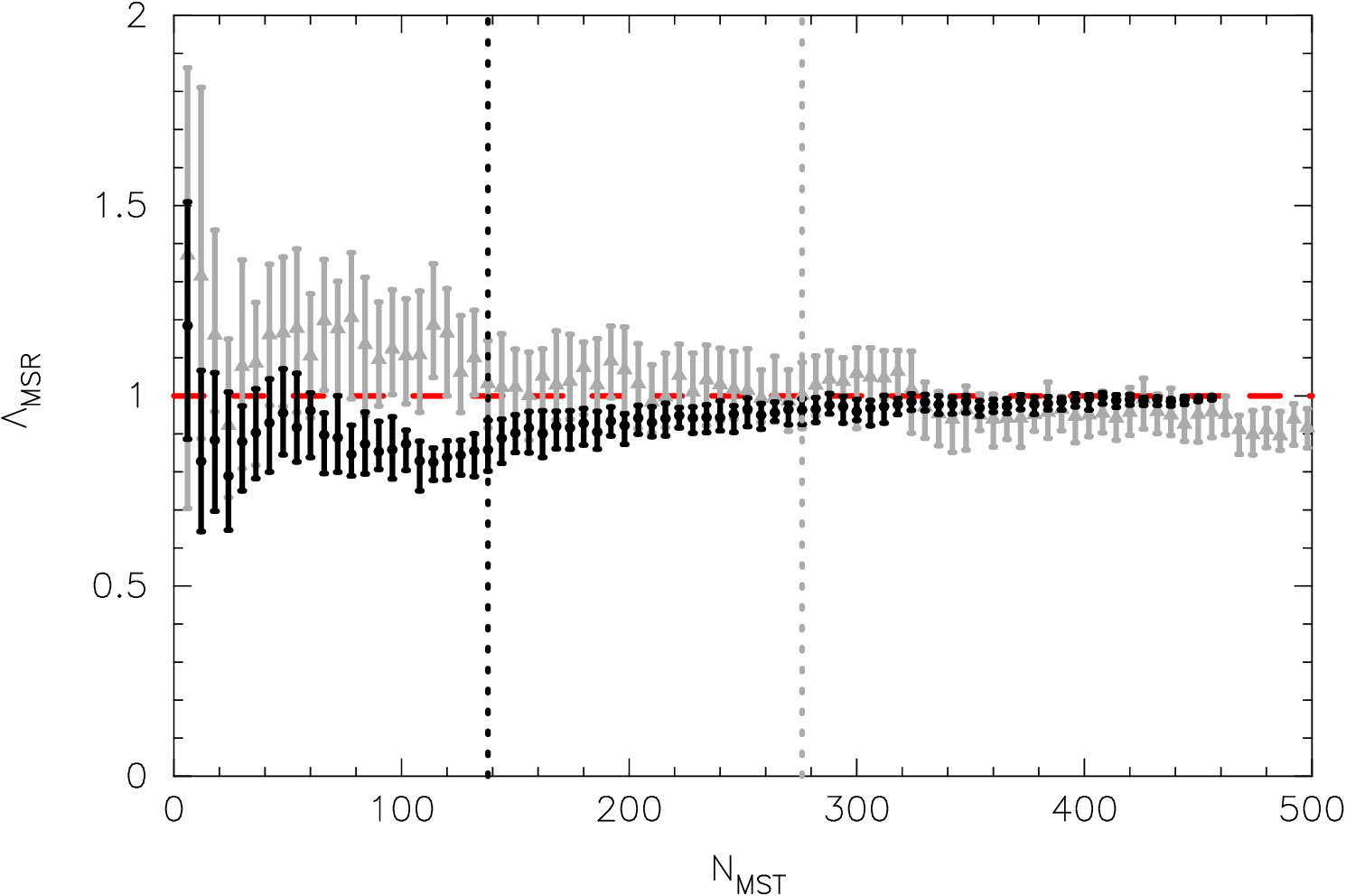}}}
\caption[bf]{The effects of non-contiguous data on the $\Lambda_{\rm MSR}$ analysis. In panel (a) we show the positions of 1000 stars in a Plummer sphere, and impose 
strips on the cluster to mimic the data in \citet{Andersen11}. In panel (b) we show the determination of $\Lambda_{\rm MSR}$ as a function of the $N_{\rm MST}$  least massive objects 
in the full sample (the grey triangular points/error bars) and for the restricted sample -- the 461 stars within the strips (the black circular points/error bars). The lefthand (black) vertical dotted line shows the 
boundary between brown dwarfs and stars for the restricted sample, and the righthand (grey) line shows the boundary location in the full sample. The red dashed line indicates $\Lambda_{\rm MSR} = 1$. }
\label{test}
  \end{center}
\end{figure*}

Given the difficulty in assessing whether any different spatial distribution of stars compared to brown dwarfs is an outcome of the star formation process, 
we revisit the data from \citet{Andersen11} to asess whether the decreasing star to brown dwarf ratio in the ONC is also echoed in the $\Lambda_{\rm MSR}$ 
and $\Sigma_{\rm LDR}$ ratio. The data from \citet{Andersen11} are not contiguous -- the coverage consists of a mosaic of `postage stamp'-like fields which appear as strips placed across the cluster, so we must assume that the observed distribution of stars and brown dwarfs 
is also representative of that in the `missing' data. 

In order to test the performance of $\Lambda_{\rm MSR}$ and $\Sigma_{\rm LDR}$ on non-contiguous data, we create a \citet{Plummer11} sphere with 1000 stars drawn randomly from the \citet{Maschberger13} IMF 
and also positioned at random. These positions are shown by the grey points in Fig.~\ref{test-a}. In Fig.~\ref{test-b} we show the $\Lambda_{\rm MSR}$ measurement as a function of $N_{\rm MST}$ for the brown dwarfs by the grey triangular points and their uncertainties. 
The location of the boundary between stars and brown dwarfs is shown by the righthand vertical dotted grey line. Whilst the calculation is quite noisy for low $N_{\rm MST}$, the values are consistent with unity. We then draw strips on the cluster and repeat the analysis, restricting the sample to the 461 stars within these strips, 
but allow MST links between stars in different strips.  The results are shown by the black circular points (and uncertainties) in Fig. ~\ref{test-b} and the location of the boundary between stars and brown dwarfs is shown by the lefthand vertical dotted black line.  Allowing MST links between the strips does 
give a small `depression' in the progression of $\Lambda_{\rm MSR}$ as a function of $N_{\rm MST}$, but the 60 lowest mass brown dwarfs have a $\Lambda_{\rm MSR}$ consistent with unity. The $\Sigma_{\rm LDR}$ value for the full sample is 1.08 (with a KS p-value 0.92), whereas the $\Sigma_{\rm LDR}$ value for the restricted sample 
is 0.79 (with KS p-value 0.16). Therefore, in both samples the ratio is not significantly different from unity. We therefore conclude that the unusual geometry of the ONC data should not affect the determination of either $\Lambda_{\rm MSR}$ or $\Sigma_{\rm LDR}$.

\begin{figure}
\begin{center}
\rotatebox{270}{\includegraphics[scale=0.35]{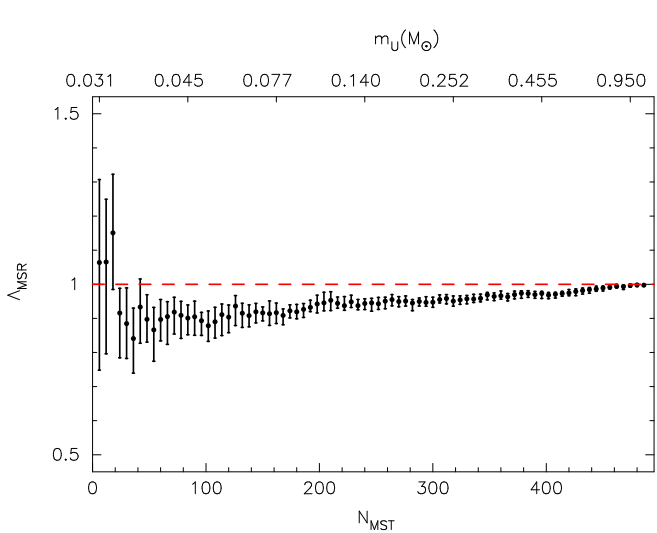}}
\end{center}
\caption[bf]{The evolution of the mass segregation ratio, $\Lambda_{\rm MSR}$, for the $N_{\rm MST}$  least massive objects in the observational sample in \citet{Andersen11}.  We indicate the 
highest mass star, $m_{\rm U}$ within the $N_{\rm MST}$.  Error bars show the 1/6 and 5/6 percentile values from the median. The dashed 
line indicates $\Lambda_{\rm MSR} = 1$,  i.e.\,\,no mass segregation.}
\label{ONC_MSR_lm}
\end{figure}

\begin{figure}
\begin{center}
\rotatebox{270}{\includegraphics[scale=0.4]{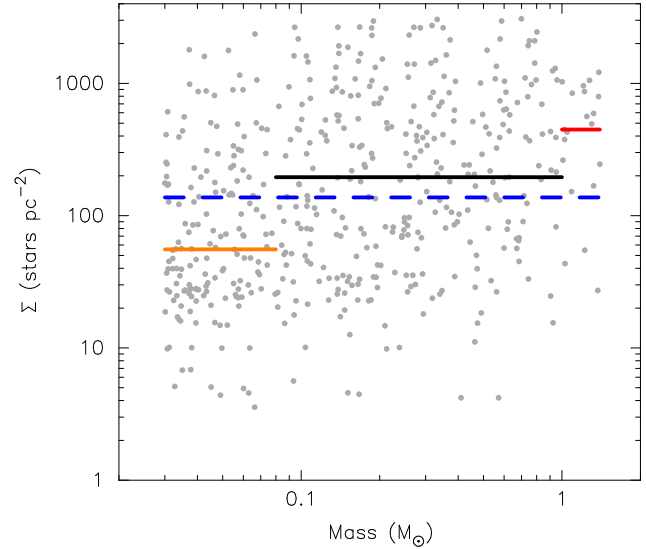}}
\end{center}
\caption[bf]{The distribution of local stellar surface density, $\Sigma$, as a function of mass, $m$, for the objects in the observational sample in \citet{Andersen11}. The median surface density for the full sample is 
shown by the dashed (blue) line. The median surface density of brown dwarfs ($0.03 \leq m/{\rm M_\odot} \leq 0.08$) is shown by the left-hand (orange) line; the median surface density of low-mass stars ($0.08 < m/{\rm M_\odot} \leq 1.0$) 
is shown by the middle (black) line and the median surface density of high-mass stars in the \citeauthor{Andersen11} observational sample ($1.0 < m/{\rm M_\odot} \leq 1.4$) is shown by the right-hand (red) line.}
\label{ONC_Sig_m}
\end{figure}

Using the data from \citet{Andersen11} we first determine $\Lambda_{\rm MSR}$ as a function of the $N_{\rm MST}$  least massive objects in the observational sample as shown in Fig.~\ref{ONC_MSR_lm}. 
The data show a marginally more spread-out spatial distribution of the brown dwarfs compared to the cluster average, although the most extreme value is 
$\Lambda_{\rm MSR} = 0.84^{+0.09}_{-0.10}$ for the 36 least massive objects, which is barely significant. $\Lambda_{\rm MSR} = 1$ (i.e.\,\,no mass segregation) is shown by the dashed line.

We then plot the local surface density $\Sigma$ against object mass $m$ in Fig.~\ref{ONC_Sig_m}, using the surface densities calculated for the whole non-contiguous sample. The median surface density for brown dwarfs is $\Sigma_{0.03 \leq m/{\rm M_\odot} \leq 0.08} = 56$\,stars\,pc$^{-2}$, shown by the horizontal orange 
line, compared to $\Sigma_{0.08 < m/{\rm M_\odot} \leq 1.0} = 196$\,stars\,pc$^{-2}$ for stars, shown by the horizontal black line ($\Sigma_{\rm LDR} = 0.29$). A KS 
test between the two distributions gives a p-value $< 10^{-7}$  that the two subsets share the same parent distribution.

We also repeated the above analysis but limited the data to objects within 1\,pc of the ONC centre and found similar results, 
suggesting that any field star contaminants in the data do not influence our analysis.

In tandem with the $\mathcal{R}_{\rm SSR}$ ratio, $\Lambda_{\rm MSR}$ and $\Sigma_{\rm LDR}$ both suggest that the spatial distribution of BDs is different to stars in the ONC. 
However, this may not necessarily be a primordial signature of star formation, as we have seen in $N$-body simulations where 6/20 clusters have a dynamical evolution that leads to spatial differences 
between stars and BDs.

\section{Conclusions}
\label{conclude} 

We have used three different diagnostics to look for differences in the spatial distributions of stars compared to brown dwarfs in $N$-body simulations of star-forming regions. We find that determining the $\mathcal{R}_{\rm ss}$ ratio 
as a function of distance from the cluster centre cannot be used on its own to draw conclusions on the spatial distribution of BDs compared to stars. In a cluster with 
a radially decreasing $\mathcal{R}_{\rm ss}$ ratio, the brown dwarfs may have a spatial distribution that is indistinguishable from stars ($\Lambda_{\rm MSR} = 1$, $\Sigma_{\rm LDR} = 1$).

Similarly, the inverse can also be true; the BDs have a significantly different spatial distribution compared to stars in that they are more spread out ($\Lambda_{\rm MSR} << 1$ and/or $\Sigma_{\rm LDR} << 1$), but the $\mathcal{R}_{\rm ss}$ ratio increases or remains constant towards the 
outskirts of the cluster.  These findings lead us to strongly advocate the use of more than one diagnostic when assessing the spatial distributions of BDs compared to stars in star-forming regions.

When applied to data from the ONC, the $\mathcal{R}_{\rm SSR}$ ratio \emph{and} $\Sigma_{\rm LDR}$ ratio -- and tentative evidence from $\Lambda_{\rm MSR}$ --  do suggest that the BDs are more spread out than stars. 
However, this dataset is spatially incomplete, and a more comprehensive survey of the ONC would be highly desirable.

Randomly distributing masses drawn from an IMF can result (in 1/20, or 5\,per cent, of simulations) in a radially decreasing $\mathcal{R}_{\rm ss}$ ratio before dynamical evolution; which may or may not be mirrored in the $\Lambda_{\rm MSR}$ and 
$\Sigma_{\rm LDR}$ measurements. Furthermore, dynamical evolution leads to significant differences between the spatial distributions of stars and BDs in more than 25\,per cent of our simulations. This implies that a large observational 
sample of regions/clusters is needed to assess whether the primordial spatial distributions of stars and BDs are different (which would suggest that their formation mechanisms are different).

\section*{Acknowledgements}

We thank the anonymous referee for their comments and suggestions, which have improved the manuscript. The simulations in this work were performed on the \texttt{BRUTUS} computing cluster at ETH Z{\"u}rich. RJP acknowledges support from the Swiss National Science Foundation (SNF). MA acknowledges support from the ANR (SEED ANR-11-CHEX-0007-01). 

\bibliographystyle{mn2e}
\bibliography{general_ref}

\label{lastpage}

\end{document}